\documentclass[%
  superscriptaddress,
  twocolumn,
  10pt,
  aps,
  pre,
  nofootinbib, 
  floatfix
]{revtex4-1}

\usepackage{graphicx,hyperref}
\usepackage{amsmath,amsfonts,amssymb}
\usepackage[utf8]{inputenc}
\usepackage{color}

\pdfoutput=1

\usepackage{newfloat}
\DeclareFloatingEnvironment[
    fileext=loa,
    listname=List of Algorithms,
    name=ALG.,
    placement=htbp,
]{algorithm}

\begin{document}
\title{Efficient limited-time reachability estimation in temporal networks}

\author{Arash Badie-Modiri}
\affiliation{Department of Computer Science, School of Science, Aalto University, FI-0007, Finland} 
\author{Márton Karsai}
\affiliation{Department of Network and Data Science, Central European University, H-1051 Budapest, Hungary}
\affiliation{Univ Lyon, ENS de Lyon, Inria, CNRS, Université Claude Bernard Lyon 1, LIP, F-69342, LYON Cedex 07, France}
\author{Mikko Kivelä}
\affiliation{Department of Computer Science, School of Science, Aalto University, FI-0007, Finland} 

\date{\today}

\begin{abstract}
Time-limited states characterise many dynamical processes on networks: disease infected individuals recover after some time, people forget news spreading on social networks, or passengers may not wait forever for a connection. These dynamics can be described as limited waiting-time processes, and they are particularly important for systems modelled as temporal networks. These processes have been studied via simulations, which is equivalent to repeatedly finding all limited-waiting time temporal paths from a source node and time. We propose a method yielding orders of magnitude more efficient way of tracking the reachability of such temporal paths. Our method gives simultaneous estimates of the in- or out-reachability (with any chosen waiting-time limit) from every possible starting point and time. It works on very large temporal networks with hundreds of millions of events on current commodity computing hardware. This opens up the possibility to analyse reachability and dynamics of spreading processes on large temporal networks in completely new ways. For example, one can now compute centralities based on global reachability for all events or can find with high probability the infected node and time, which would lead to the largest epidemic outbreak.
\end{abstract}

\maketitle

\section{Introduction}

The topology of networks laying behind complex systems is crucial for any dynamical processes taking place on them~\cite{barrat2008dynamical}. This realisation provided new perspectives in understanding various phenomena, such as spreading of disease~\cite{van2011gleamviz} and social dynamics~\cite{castellano2009statistical}. In addition to the topology, it has later become evident that the time-varying nature of these connections also has a large effect on the unfolding of spreading processes~\cite{masuda2017temporal} and many other dynamical phenomena~\cite{Gauvin2018randomized}. This was one of the main realisations leading to the emergence of the field of temporal networks, which studies structures where links are not static but active only at some specific times~\cite{holme2012temporal,holme2015modern}. The timing of connections has both uncovered interesting phenomena never seen before and created new types of computational problems to the analysis of network data and models.

In static networks, the possible routes for any dynamics to evolve are determined by topological paths. Paths can also be defined for temporal networks, but there are two main fundamental differences. First, the paths need to be time respecting such that the consecutive links are activated in the correct order~\cite{Holme2005Network}. Second, the time between activations is often limited. This is because many of the processes are characterised by time-limited states and finite memory, e.g., in case of spreading processes where they appear as the limited lifetime of a spreading agent. The maximum acceptable transfer time in a transportation network~\cite{guo2011assessing,alessandretti2016user} or in a gossip protocol~\cite{braginsky2002rumor}, as well as the finite infectious period of an individual in case of disease spreading~\cite{pastor2015epidemic} are all good examples of such dynamics. These processes can only spread through time-respecting paths where consecutive connections take place within some limited time $\delta t$.

The detection of temporal paths and the connectivity they provide is fundamental to understanding dynamics on and characteristics of the networks, but it cannot directly rely on the methodologies developed for static structures. Instead, new methods need to be developed, and this work is still at its infancy compared to static networks. For example, temporal connectivity and related measures are routinely being computed using breadth-first search type of algorithms. This is similar to the approach of finding connected components in static graphs in the early studies on percolation phenomena on lattices~\cite{Leath1976Cluster}. Major improvements to these early algorithms, such as the Newman-Ziff algorithm~\cite{Newman2001Fast}, made it possible to analyse large network structures in an unprecedented way and opened the path to the understanding of the connectivity of networks we have today.

An elegant way to overcome difficulties in temporal networks is to transform the temporal problems into static problems, which we know how to solve efficiently. To do so, we need a representation, which maps temporal networks to a static structure on which we can then apply static network methods. Weighted temporal event graphs have been recently suggested as one such solution~\cite{kivela2018mapping, mellor2017temporal}. They provide a representation of temporal networks as static directed acyclic graphs (DAGs), which contains all the information on the temporal paths. They can be interpreted as temporal line graphs, where events are nodes and if they are adjacent, they are connected by a link directed according to the arrow of time. Such links can form longer $\delta t$-constrained path, representing the ways a limited waiting-time process can spread in the structure. This representation allows us to design efficient algorithms to measure temporal centrality or connectivity in time-varying networks, while exploiting tools and theories developed for static graphs and directed acyclic graphs. 

A particular way of using the weighted event graphs is to use the Newman-Ziff algorithm~\cite{Newman2001Fast} to measure the size of the weakly-connected component when increasing the $\delta t$ value~\cite{kivela2018mapping}. This allows extremely fast sweeps of $\delta t$ values where the size of the weakly-connected components can be measured for each value. However, weakly connected components only give an upper estimate for the outcome of any potential global phenomena. On the other hand, more precise indicators of connectivity and influence, like in- and out-components, are difficult to measure with current conventional techniques.

Here we take a complementary approach to the Newman-Ziff algorithm and develop a method to make accurate estimates of the sizes of source and influence set of every single event in a temporal network, given an arbitrary $\delta t$. We rely on the DAG character of the event graph representation, which allows us to convert our temporal reachability problem to a DAG reachability problem, a.k.a., the graph-theoretical challenge to estimate transitive closure sizes~\cite{Cohen1996Size}. Relying on already developed probabilistic counting methods~\cite{heule2013hyperloglog}, we can devise an algorithm, which estimates the global reachability for each event even in extremely large temporal networks with hundreds of millions of events. Further, using this approach, we can effectively identify with high probability events with the largest out- (and in-) components in massive temporal networks.

To introduce and demonstrate our method, first in Section~\ref{sec:defDAG} we define the basic formal concepts of event graphs. In Section~\ref{sec:algo} we describe our algorithmic solutions and use them in Sections~\ref{sec:rn} and \ref{sec:realn} to estimate out-component sizes of events in random and real-world networks. Analysed networks include large-scale temporal structures such as mobile phone communication and transportation networks. Note that the implementation of the algorithms described in Section~\ref{ch:methods} and the Appendices are publicly available~\cite{badie2019implementation}.

\section{Methods}
\label{ch:methods}

\subsection{Definitions and measures}
\label{sec:defDAG}

\subsubsection{Temporal networks, adjacency, temporal paths, and reachability}
Temporal networks can be formally defined in various ways~\cite{holme2012temporal}. They build up from time-varying interaction events, which can be directed or undirected, appear with duration or delay, and can be between two or more nodes. In turn, events induce temporal paths, whose structure critically depends on the event characteristics. To capture all of this complexity, we introduce methods using a slightly more general definition of temporal networks than usual, which can be easily made more specific depending on the features of the actual temporal network.

We define a temporal network as a tuple $G=(V_G,E_G,T)$ of a finite set of nodes $V_G$, a finite set of events $E_G$, and an observation window $T$. An event $e \in E_G$ is defined as $e=({\bf u},{\bf v},t,\tau)$, where ${\bf u },{\bf v} \subseteq V_G$ are the source and target node sets of the event, $t$ is the time at which the event starts, and $\tau$ is its delay or duration\footnote{Note that nodes here formally appear as set of nodes ${\bf u}, {\bf v}\subseteq V_G$ to compile with possible hyper events in the representation, however in case we assume that only a pair of nodes can participate in an event, the event definition relaxes to the usual case where $e=({u},{v},t,\tau)$, where $u$ and $v$ are single nodes from $V_G$.}. Here we assume that the source and target event sets are relatively small with a constant size, not depending on the length of the temporal data, which is usually the case in real temporal networks.

To capture possible information flow~\cite{kivela2012multiscale}, potential causal relationships~\cite{karsai2011small}, and mesoscopic motifs in temporal networks~\cite{kovanen2013temporal}, we can define the \emph{adjacency} relation between pairs of events. We say that two events $e_i,e_j\in E_G$ are \emph{adjacent}, $e_i \rightarrow e_j$, if they have at least one node in common in their target and source sets, $\bf{v}_i \cap \bf{u}_j \neq \emptyset$, and they are consecutive: the second event $e_j$, at $t_j>t_i$ cannot start before the first event $e_i$ ends, thus the time difference between the two events must be $\Delta t(e_i,e_j)=t_j - t_i - \tau_i >0$. In addition, we can constrain events to be $\delta t$-\emph{adjacent}, $e_i \xrightarrow{\delta t} e_j$, thus being related only if they happen within a time distance $\delta t$, i.e., $\Delta t(e_i,e_j) \leq \delta t$.

Unlike in static networks, in temporal structures information can pass between nodes only at the time and direction of interactions. Thus to study any dynamical process on temporal networks, we first need to define how information can be propagated through a sequence of events. We define a temporal path (also called a time-respecting path) as an alternating sequence
\begin{equation} \label{def:tpath}
P=[v_1, e_1, v_2, e_2, \dots, e_n,v_{n+1}]\,,
\end{equation}
of $v_i \in V_G$ nodes, $e_i \in E_G$ events, which must be adjacent if they are consecutive in the sequence. In contrary to static paths, a temporal path is not permanent but depends on the time and the source node of the first interaction. Moreover, in a temporal path consecutive events need to be adjacent: they need to happen in correct temporal order while taking account their duration and direction as well. In addition, we can constrain consecutive events to be $\delta t$-adjacent, to capture processes with a maximum allowed transfer time. Taking these possible restrictions, we can already code some characters of the dynamical process in the representation of the underlying temporal network. In the following, we often use an example a simplification of this general description by assuming instantaneous, undirected, and dyadic interactions with only two interacting nodes~\cite{karsai2011small}. This gives us a network $G^\prime$ with an event set $E_{G^\prime} \subset V\times V \times [0,T]$ and an event defined as $(u,v,t) \in E_{G^\prime}$\footnote{Note that in case of undirected dyadic events we can store an event in a more general form as $(\{u,v \},\{u,v \},t,0 )$ but in practice in our algorithms there is no reason to explicitly store both source and target sets and zeros for the delays.}.

Temporal paths code \emph{reachability} in a temporal network, i.e., whether a node at a given time can or cannot influence another node in an upcoming time step. Considering all outgoing (or incoming) temporal paths starting from (resp. ending at) a given node at a given time, one can obtain its influence (resp. source) set. This out-component (resp. in-component) can be computed as the union of the nodes in the time respecting paths starting from (ending at) a given node. The out-component determines the possible routes information, epidemics, rumour or influence can travel after initiated from a given node at a given time. This may give us the potentially infected set of patients in an epidemic, or the influenced set of people of a political campaign. However, the solution of the reachability problem is computationally expensive even for small structures~\cite{Holme2005Network}. For larger temporal networks the only feasible solution has been to sample initial source node-time pairs and compute their influence sets using a breadth-first search algorithm~\cite{Holme2005Network,cormen2009introduction}. This approach, although very expensive, has already provided some insight about the average reachability of temporal networks and its relation to various network features~\cite{karsai2011small, Gauvin2018randomized}.

\subsubsection{Weighted temporal event graphs}

A recently introduced higher-order representation of temporal networks, called \emph{temporal event graphs}, provides effective solutions to many computational problems related to temporal network connectivity~\cite{kivela2018mapping} and other purposes~\cite{mellor2017temporal}. Given a temporal network $G=(V_G, E_G, T)$, the temporal event graph representation is formally defined as a weighted graph $D = (E_G, A_E, \Delta t)$. The nodes of $D$ are the events of $G$ and edges are drawn between adjacent events. The direction of  every edge is in the arrow of the time and the weight is defined as the time difference of the two events incident to the edge, $\Delta t(e_i, e_j)$. Going forward in time, the direct successors of event $e \in E_G$ are the set of events connected by outgoing edges from $e$. Going backwards in time, direct predecessors of an event $e$ are the set of all events where there is a directed edge from that event to $e$. Event graphs can be regarded as a temporal line graph representation, capturing higher-order relationships between events.

Since adjacency is defined between non-simultaneous events and directed by time, temporal event graphs appear as weighted directed acyclic graphs (DAG). They are static representations of temporal networks, which can be analysed by the full spectrum of tools and methods developed for static graph. Further, they allow one to use concepts of static centrality and similarity measures do develop similar concepts in temporal networks. As their most important feature, they appear as a static superposition of all temporal paths present in the original temporal network. In other words, $\mathcal{P}_e^G = \mathcal{P}_v^D$, where $\mathcal{P}_e^G$ is the set of event sequences in all temporal paths in $G$, and $\mathcal{P}_v^D$ is the set of node sequences in all the paths in $D$. 
While every temporal path in $G$ corresponds to a unique path in $D$, there can be redundancy in the other direction: multiple temporal paths in $G$ could correspond to a single event path in $D$. This is due to the multiple temporal paths using the same sequence of events, but a different sequence of nodes. It is easy to construct such sets of paths with events that have multiple source and target nodes. In the case of dyadic interactions, such redundancies are very minor (or non-existent). In any case, these multiplicities do not have any effect on the reachability.

\begin{figure}
    \centering
    \includegraphics[width=0.95\linewidth]{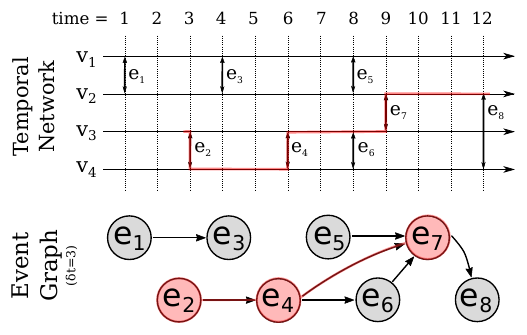}
    \caption{\label{fig:temporal-net} An undirected temporal network and its corresponding static event graph. Two events $e_i$ and $e_j$ on temporal network are adjacent if they share at least one node and $\Delta t(e_i,e_j) \leq \delta t$ where in this instance we have chosen a value of $\delta t = 3$ for the purpose of illustration. One valid temporal path is highlighted red on the temporal network and on the event graph.}
\end{figure}

\subsubsection{Component definitions}\label{sec:comp-def}
Components can be defined in various ways in an event graph. As it is a directed graph, one can identify in- and out-components and also weakly connected components. Since event graphs are directed acyclic graphs, strongly connected components larger than one node do not exist.

More precisely, the \emph{out-component} of an event (also called the root event) in a static event graph 
is defined as the maximum set of other future events that can be reached by any temporal path starting from the root event. In case of an epidemic spreading with initial infection taking place at the root event, this is the set of temporal contacts (and nodes) which potentially propagate the disease. We define the \emph{maximum out-component} of an event graph as its largest out-component, giving us the largest possible effect/outbreak ever observable in the network. Equivalently, the \emph{in-component} of an event is formed by the incoming temporal paths and it is defined as the set of earlier events (and nodes), which can influence the actual pair of interacting nodes up to the actual time. The definition of \emph{weakly connected components} is less restrictive as they include any events, which are connected via temporal paths irrespective of the direction of their adjacency.

Among all these component types, in the following we are mostly going to focus on the precise identification of \emph{out-components}, and we will explain how our methodology can be generalised to identify in-components as well. The end goal for our algorithm is to rapidly determine the sizes of these components.

Temporal graphs provide further ways to define connectivity~\cite{kivela2018mapping}. Beyond connected events in the components of $D$, one can look for the set of original network nodes from $V_G$ involved in such components. Since a network node can appear in multiple times in an event graph component, this is an alternative way to measure the influence of an event by counting the total number of network nodes involved in the corresponding event graph out-component. Event graph components have also temporal dimensions, thus their connectivity can be also measured in terms of the time span between their first and last events. This compared to the $T$ total observation time tells whether a component has only a local temporal effect or it percolates in time and bridges information over a longer course of observation.

\begin{figure}
    \centering
    \includegraphics[width=0.7\linewidth]{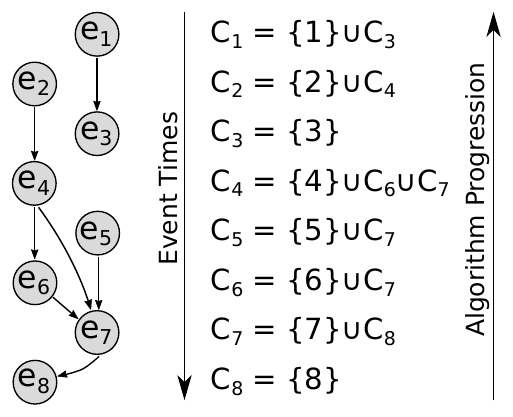}
    \caption{\label{fig:example-dag} Static event graph representation of a temporal network. Weakly connected components are $\{e_1, e_3\}$ and $\{e_2, e_4, e_5, e_6, e_7, e_8\}$. Out-component set of the event $e_2$ is $\{e_2, e_4, e_6, e_7, e_8\}$. Note that event $e_5$ is in the same weakly connected component as $e_2$ but it is not a member of its out-component set. The algorithm finds the out-component by going backwards through a topological ordering of events (i.e.~reverse time order), at each step, the out-component of each event is calculated by getting the union of the out-component sets of all the events from the set of events in its out-edges plus the event itself. Since the ordering is reverse topological order, all the events in the out-edge set will already have their out-components calculated.}
\end{figure}

\subsection{Scalable algorithms for in- and out-component size estimation}
\label{sec:algo}

The out-component of an event $e_i \in E_G$ (which is a node in $D$) can be calculated in several ways. As we have mentioned, one potential solution is to start a breath-first-search process from one of the nodes involved in $e_i$ by using the upcoming events in $G$. Another solution would be to compute the direct successor set recursively using the algorithm explained in Appendix~\ref{sec:redWTEGs}. However,
calculating the sizes of out-components even for a small fraction of all events is not feasible with any of these solutions for large temporal networks, as their complexity scales badly with the number of events $|E_G|$. Here we propose an alternative solution based on a probabilistic approach to estimate the size of the largest out-component to arbitrary precision and to identify its root event in any temporal network, even with extremely large sizes. 

\subsubsection{Probabilistic method for estimating out-component sizes}\label{sec:out-comp}
Our main goal is to obtain the out-component size for each node in $D$. But for a more concise presentation, first, we define an algorithm, which exactly provides out-components (i.e., the reachable sets) for each node in $D$. Our out-component size estimation algorithm is subsequently defined by changing the data structure containing these sets to a probabilistic counting data structure~\cite{Cohen1996Size,flajolet2007hyperloglog,heule2013hyperloglog}.

Our solution is similar to the commonly used algorithm for computing all subtree sizes, where starting from leaf nodes, the size of each subtree is given by the sum of its subtree sizes plus one. We tailor this idea specifically for DAG structures. This algorithm reuses the already computed out components for direct successors to calculate the out component of each node in a directed acyclic graph. To explain the algorithm we consider separately the nodes with zero and non-zero out-degrees $k_{out}$: The out component 
of any leaf node $i$ (i.e.~node with $k_{out}=0$) is trivial as it contains only itself $C_i=\{ i \}$. For the other nodes ($k_{out}>0$) the out component 
can be built as $C_i = \{ i \} \cup \bigcup_{(i,j) \in A_G} C_j$  where $A_G$ is the
set of edges in the event graph $D$.We compute the out components $C_i$ by going through the nodes of $D$ (events of the temporal network) in reverse topological order, for example, reverse temporal order starting from the event with the largest timestamp backwards. This ensures that we already know the out-components of direct successors of each event we encounter. The algorithm is illustrated in Figure~\ref{fig:example-dag} and described in detail in the Algorithm~\ref{algo:out-component-size}.

\begin{algorithm}
    \centering
    \includegraphics[width=0.90\linewidth]{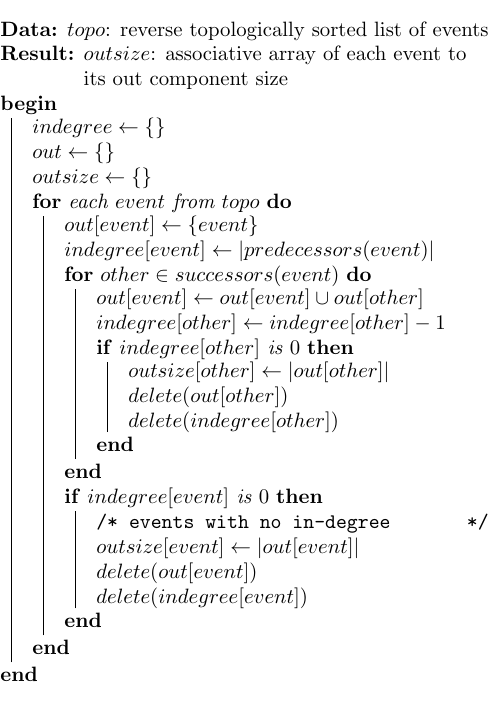}
    \caption{\label{algo:out-component-size} Calculating out-component sizes of events from the static event graph representation described in Appendix~\ref{sec:event-graph-rep}. $Successor(e)$ (and $Predecessor(e)$) return set of direct successors (and predecessors) of event $e$ as described in Appendix~\ref{sec:event-graph-rep} and Algorithm~\ref{algo:succ}. Associative array $out$ is used to keep the memory-intensive set representation of out-components of events in memory up until the moment when there would be no references to the out-component of that event, when they are deleted from $out$ and the cardinality of their out-component set is added to $outsize$.}
\end{algorithm}

This algorithm only goes through each event once and performs number of union operations equal to the number of links in the event graph, $|A_G|$. However,
the average out-component size can be directly proportional to the number of events in well-connected networks. That is, the out components can grow rapidly when the network size grows. This makes the algorithm to scale badly both in memory and computational time due to the cost of union operations on increasingly large sets.

The root of the performance problem is that we store the actual reachability sets when we only need their sizes. The solution to this problem is to find a data structure to replace the sets $C_i$ with another data structure $\hat{C}_i$, which has a constant size and constant time union operator $\hat{C}_i \cup \hat{C}_j$ and can return an estimate for the set size $|\hat{C}_i|$ (again in constant time). With this data structure, the scaling of the algorithm becomes $\mathcal{O}(|E_G|log(|E_G|)+|A_G|)$, which is much preferable to the breadth-first search approach with $\mathcal{O}(|E_G||A_G|)$ complexity. Probabilistic counting methods described next give access to exactly this type of data structures.

The method described above works equally well if we want to measure the sizes of the components in terms of nodes of the temporal network $G$. In this case, the reachable sets would be populated with the nodes of the events instead of the events themselves. If the sizes are measured in lifetimes, i.e., the time between the first and the last event in the component the algorithm can be made even more simple. In this case, instead of saving the full reachable set of nodes, it is enough to save the largest timestamp of all of the event. That is, the set $C_i$ is replaced with a timestamp $T_i$, which is initially set to $t_i$ for any event $e_i$ appearing as a leaf node in $D$, and the union operator is replaced with the maximum operator. 

Note that although here we discuss the computation of the out-component sizes, in-components can be calculated with the same algorithm by reversing the direction of the links in $D$ and the order at which the nodes in $D$ are traversed. In practice the reversion of the link direction can be obtained by replacing calls to $Successors(e)$ function with $Predecessors(e)$ and vice versa in Algorithms~\ref{algo:out-component-size} and \ref{algo:bfs}.

\subsubsection{Probabilistic cardinality estimator}

For Algorithm~\ref{algo:out-component-size} to run on large real-world networks we need to ensure that the time complexity of the union and the cardinality operators and also the space complexity of the set implementation do not grow linearly as a function of the cardinality of the set. This is not the case for 
implementations which exactly keep track of the out component sets for example using sorted vectors or hash maps. However, in order to estimate out-component sizes it is not necessary to query the sets for their members, but only to insert, merge and query the size of each set. We use a data structure implementing a variation of the HyperLogLog probabilistic cardinality estimator algorithm~\cite{flajolet2007hyperloglog}, which is computationally efficient for the three required operations. HyperLogLog was conceived as a method of estimating the cardinality of massive multisets, usually in the form of streams, given a constant amount of memory.

The basic idea of the algorithm is to use randomisation, in form of passing the input through a hash function, and only save the maximum number of leading zeros in the binary representation of the hashed values of the multiset. A cardinality estimation is then made by counting the number of leading zeros. Due to the uniform distribution of the output of hash functions suited for this algorithm, if the maximum number of observed leading zeros is $\varrho-1$ then a good estimation of the cardinality would be $2^\varrho$. Alone, the above-described estimators are extremely crude, but the algorithm works by combining many such estimators via a process of \emph{stochastic averaging}. Based on the hash value the algorithm splits the input stream into $m$ substreams while keeping track of the maximum number of leading zeros in each substream. Subsequently, it averages the observables using their harmonic mean, which ensures that variability of the estimation is kept in check~\cite{flajolet2007hyperloglog}.

\begin{figure}
    \centering
    \includegraphics[width=0.49\linewidth]{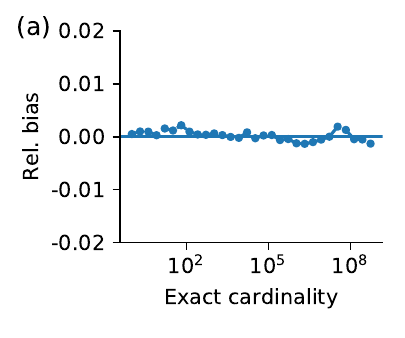}%
    \includegraphics[width=0.49\linewidth]{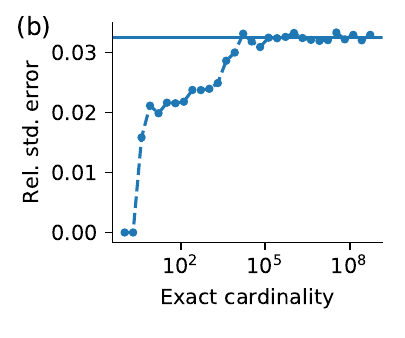}
    \includegraphics[width=0.49\linewidth]{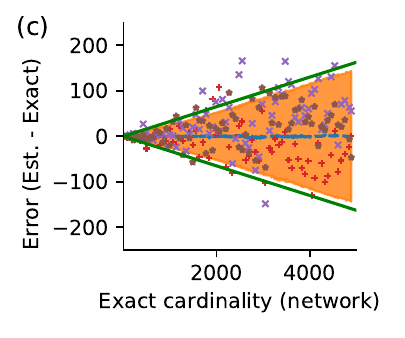}%
    \includegraphics[width=0.49\linewidth]{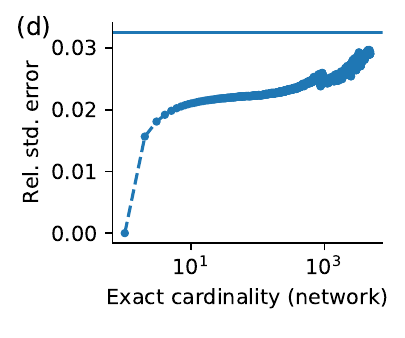}
    \caption{(a) Estimates of mean relative bias values, (b) relative standard error of our implementation of HyperLogLog cardinality estimation algorithm with $m = 2^{10}$ registers and random multisets. Relative standard error is calculated by dividing standard deviation of estimated cardinality over a set of $1\,000$ estimators with different seeds divided by true values of cardinalities. Relative bias is measured as difference of the estimated cardinalities from the true cardinality divided by true cardinality averaged over $1\,000$ estimators. Panel (c) displays mean (blue dashed line) and the standard error of the mean (orange band) of $1\,000$ independent estimations of out-component size estimations on a realisation of 1024-node random temporal network. The markers $\ast$, $+$ and $\times$ denote the estimation errors for a subset of individual out-component sizes for three estimators out of the total $1\,000$. Panel (d) shows relative standard error of estimation, similar to (b), but based on $1\,000$ independent out-component size estimation on the same network as (c). Solid lines in panels (b), (c) and (d) indicate theoretical accuracy of HyperLogLog algorithm with $2^{10}$ registers, which is $1.04/\sqrt{m} = 1.04/\sqrt{2^{10}} = 0.0325$ relative to true cardinality~\cite{flajolet2007hyperloglog}. For a more in-depth study of different variations of HyperLogLog and the role and reasoning about bias-estimation see Ref.~\cite{heule2013hyperloglog}.}
    \label{fig:hll-error-bias}
\end{figure}

We made several choices in our implementation of the algorithm, with some described in more details in the definition of HyperLogLog++ algorithm~\cite{heule2013hyperloglog}. In particular, the following modifications were borrowed from HyperLogLog++:
(a) We used a 64-bit hash function, as opposed to original 32-bit, to compensate for the collision of hash values for multisets with large cardinalities.
(b) Empirical bias correction was performed as introduced in~\cite{heule2013hyperloglog}.
(c) To improve performance characteristics and simplify error analysis, we did not use a separate sparse representation. Fig.~\ref{fig:hll-error-bias} shows the relative accuracy and bias values for the HyperLogLog cardinality estimator. The difference in the scale of bias and accuracy indicates that the bias estimation reduced the bias and stopped its growth as the cardinality grows, to a degree where it plays an insignificant role in the total inaccuracy of the estimator.

The relative error in the size estimates can be made arbitrarily small by increasing the number of registers $m$. Estimations of cardinality of a multiset $S$ is expected to have a Gaussian distribution, due to averaging and the central limit theorem, with a mean of $|S|$ and a standard deviation of $\frac{1.04 |S|}{\sqrt{m}}$ (for $m > 128$)~\cite{flajolet2007hyperloglog}. HyperLogLog needs at most 6 bits ($log_2(64)$) per register to store the number of leading zeros in the output of the 64-bit hash function but for ease of use, we elected to assign a full 8-bit byte for each register. A HyperLogLog counter has been reported to be able to estimate cardinalities well beyond 1 billion, limited by raising collision probability as approaching to the limits of a 64-bit hash function~\cite{heule2013hyperloglog}. As an example, a counter with $m=2^{10}$ registers would have a constant size of one kibibyte and a relative accuracy (corresponding to standard deviation of the distribution of estimates as a fraction of actual cardinality) of $0.0325$. While inserting an item in the HyperLogLog estimator requires a constant number of operations with respect to cardinality or number of registers, the estimation operation requires linear operations with respect to the number of registers.

For a specific relative error rate, the memory and time-scaling of the probabilistic counter are constant and do not depend on the input network size. In practice, the constants involved are relatively large. For this reason, we only keep track of the cardinality estimator data structures for nodes that do still have unprocessed predecessors. This significantly reduces the memory requirements when running the algorithm on real data (see Section~\ref{sec:data}).

\subsubsection{Finding the event with largest out-component}

The above-described algorithm finds accurate estimates for the out-component sizes of nodes in a DAG. However, it can be further developed to design a probabilistic estimation method to find the event with the maximum out-component size with a highly adjustable probability. This is possible by complementing the estimates with breadth-first search. That is, starting from the event with largest estimated out-component size, we perform consecutive breadth-first search operations, finding exact out-component size either identifying it as the new largest out-component size or ruling it out as such. This is repeated until the probability that any of the estimated (non-exact) out-component sizes being larger than the largest exact out-component size is smaller than some predefined probability threshold.

Let's assume that the out-component size estimation process provides an $\hat{s}_e$ out-component size for the event $e$. We can calculate the probability distribution of the actual out-component size of that event $s_e$ based on the extended form of Bayes' theorem:
\begin{equation}\label{eq:extended-bayes}
    P(s_e | \hat{s}_e) = \frac{P(\hat{s}_e | s_e) P(s_e)}{\sum_{i=1}^\infty P(\hat{s}_e | i) P(i)} \,,
\end{equation}
where $P(s_e | \hat{s}_e)$ is the probability of the actual size being $s_e$ when the estimate $\hat{s}_e$ is observed, $P(\hat{s}_e | s_e)$ is the probability to estimate the size of a multiset with cardinality $s_e$ as $\hat{s}_e$, and $P(s_e)$ is the probability that any multiset would have a cardinality of $s_e$. The term $P(\hat{s}_e | s_e)$ can be approximated by a probability density function of a Gaussian distribution with a mean of $s_e$ and standard deviation of $s_e \frac{1.04}{\sqrt{m}}$ for $m > 128$, where $m$ is the number of registers of the probabilistic counter~\cite{flajolet2007hyperloglog}. Assuming a uniform prior~\footnote{The actual distributions of the component sizes will be biased towards small components especially for the regions of $\delta t$ which are of most interest. Prior with more probability mass on the large values will mean that our estimate on the number of breadth-first search operations we need to perform to achieve the desired accuracy gets larger. That is, the uniform prior is likely to be an overly cautious option as a prior for the component sizes.} for cardinality of multisets, Eq.~\ref{eq:extended-bayes} simplifies to:
\begin{equation}\label{eq:conditional-size-prob}
    P(s_e | \hat{s}_e) = \frac{P(\hat{s}_e | s_e)}{\sum_{i=1}^\infty P(\hat{s}_e | i)}\,.
\end{equation}

Assume we have estimated in- or out-component sizes of all the events as $\{\hat{s}_1, \hat{s}_2, \ldots\}$. Without loss of generality, we take that $\hat{s}_1$ is the largest estimate (i.e.~$\forall_{e \in E_G} \hat{s}_1 \geq \hat{s}_e$). If the actual in- or out-component size corresponding to event 1 is measured using the exact algorithm described in Appendix~\ref{sec:bfs} as $s_1$, the probability that $s_1$ would be the largest in- or out-component size of the whole network can be expressed as:
\begin{equation}\label{eq:conditional-population}
    P(\forall_{e \in E_G} s_1 \geq s_e | \hat{s}_1) = \prod_{e \in E_G \setminus \{1\}}{P(s_1 \geq s_e | \hat{s}_e)}\,,
\end{equation}
where given Eq.~\ref{eq:conditional-size-prob}, $P(x \geq s_e | \hat{s}_e)$ can be written as:
\begin{equation}\label{eq:noneq-prob}
    P(x \geq s_e | \hat{s_e}) = \frac{\sum_{i = x}^{\infty}{P(\hat{s}_e | i)}}{\sum_{i=1}^\infty P(\hat{s}_e | i)}\,.
\end{equation}

Along with a large enough number of registers, this can increase the probability of finding the absolute largest in- or out-component at any desirable level by removing estimates one by one through calculating exact in- or out-component sizes with the breadth-first search algorithm. It is also possible to use this technique for finding the largest out-component size to a specific number of significant figures.

\section{Applications}

To demonstrate the use of our method we first apply it on simulated (Section~\ref{sec:random}) and subsequently on empirical temporal networks (Section~\ref{sec:data}). As it comes, we focus on the computation of out-components, but in-components could also be obtained with the same method.

\subsection{Random networks}\label{sec:random}
\label{sec:rn}

For the demonstration of our methodology, we use one of the simplest temporal network model, which assumes that both the structure and the link dynamics are completely random and uncorrelated~\cite{kivela2018mapping}. More specifically, our model network is built on a static structure generated as an Erdős–Rényi random graph with $n$ nodes and $k$ average degree. Each link has an interaction dynamics modelled as a Poisson process with a rate parameter $\alpha=1$ for an observation window $T$. Thus, events on links follow each other with exponentially distributed inter-event times.
\begin{figure}[ht!]
    \centering
    \includegraphics[width=0.49\linewidth]{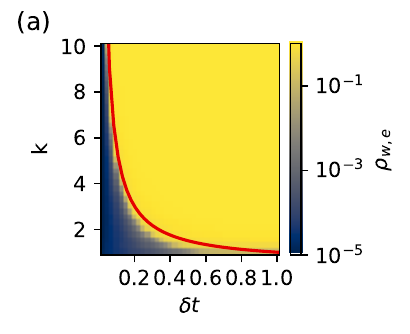}%
    \includegraphics[width=0.49\linewidth]{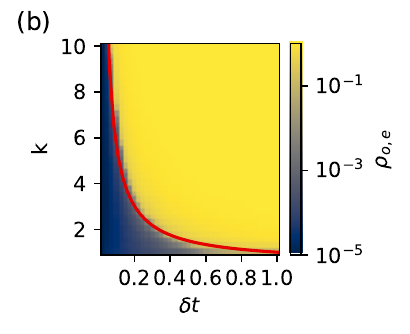}
    \includegraphics[width=0.49\linewidth]{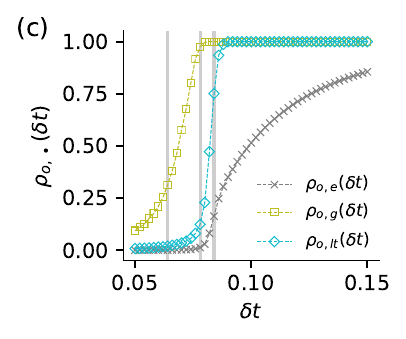}%
    \includegraphics[width=0.49\linewidth]{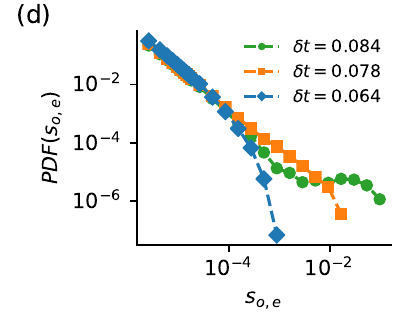}

    \caption{\label{fig:random} 
    Component sizes for a random temporal network model with an Erdős–Rényi static base and events created on each link with a Poisson process. Maximum (a) weakly-connected and (b) out-component sizes measured in events as a function of average degree and $\delta t$. Solid line indicates the analytically estimated critical point for percolation of out-components~\cite{kivela2018mapping}. (c) The growth of the maximum out-component for a fixed average degree $k=9$ for component sizes measured in events ($\rho_{o,e}$), temporal network nodes ($\rho_{o,g}$) and the lifetime of the component ($\rho_{o,lt}$). (d) Distributions of out-component sizes measured in events for networks with average degree $k=9$ at three different values of $\delta t$. These $\delta t$ values are marked with vertical lines in panel (c). Networks have $|V|=1024$ nodes evolving for $T=512$ for heatmaps (a and b) and $T=128$ time units for panel (c). Events were generated through a Poisson process with expected inter-event time of $\alpha=1$ time units. Each point is averaged over 10 realisations for heatmaps (a and b) and 50 realisations for panel (c) with $2^{14}$ registers for each HyperLogLog counter implying a relative standard error of $8.1\times10^{-3}$.}
\end{figure}

It has been shown earlier~\cite{kivela2018mapping} that by varying the $k$ average degree and the $\delta t$ event adjacency parameter the event graph goes through a percolation phase transition between a disconnected and a connected phase. If $\delta t$ is small or the underlying network is disconnected ($k_{out}<1$), only short temporal paths can evolve between small components of connected nodes, thus the potential sizes of DAG components are very limited. However, on a connected structure ($k_{out}>1$), by increasing $\delta t$, more and more events become $\delta t$-adjacent, this way forming longer paths and potentially larger event graph components. At a critical $\delta t$ the event graph goes through a directed-percolation-like phase transition, with an emerging giant connected component, which connects the majority of events via valid $\delta t$-connected time respecting paths. This transition has been observed earlier~\cite{kivela2018mapping} via the measurement of the largest weakly connected component of the temporal event graph, as it is demonstrated in Fig.~\ref{fig:random}a. The critical point can be approximated via a simple analytic function $\delta t_c=1/(\alpha(2k-1))$ (solid line in Fig.~\ref{fig:random}a) or via the scaling of different thermodynamic properties of the system~\cite{kivela2018mapping}. Although the analytic and simulated critical points match relatively well each other (see Fig.~\ref{fig:random}a), discrepancies between them are due to (i) the analytic solution being an approximation only underestimating the critical point and (ii) weakly connected components providing only an upper limit for the actual largest out-component sizes. However, comparing the analytic curve to the estimated largest out-component sizes we find a significantly better match, as it is shown in Fig.~\ref{fig:random}b. 

Just like in case of the weakly connected components in~\cite{kivela2018mapping}, the out-components sizes can be measured in three different ways: in terms of the number of events, the number of temporal network nodes, and in terms of the time between the first and last even in the component (i.e.~the lifetime). As discussed in Section~\ref{sec:out-comp} the algorithm presented here is easily adaptable to calculating the sizes of components in the temporal network nodes and even simpler algorithm can be used for the lifetimes. The results of these calculations for a single average degree value are shown in Fig.~\ref{fig:random}c. Further, the algorithm produces the out-component sizes for all events in the network, which allows us to study their size distribution. These distributions are shown in Fig.~\ref{fig:random}d for three $\delta t$ values around the value at which the largest component size becomes comparable to the system size. If these distributions would have been produced by sampling events and performing breadth-first search operations, the three distributions with different $\delta t$ values would have looked almost identical with $~10^4$ breadth-first search operations and difference in the tails would only become visible with an expected number of around $10^5$ to $10^6$ breadth-first search operations, which would have been comparable in terms of runtime to performing breadth-first search operations from all events.

\subsection{Real networks}\label{sec:data}
\label{sec:realn}

To benchmark the performance of the algorithm we measured reachability values of a set of real-world networks. (a) A mobile call network~\cite{karsai2011small} of 325 million events of over 5 million customers over a period of 120 days; (b) 258 million Twitter interactions~\cite{yang2011patterns} of over 12 million users over a period of more than 200 days; (c) air transport network of United States~\cite{bts2017air} with 180\,112 flights between 279 airports; and (d) public transportation network of Helsinki~\cite{kujala2018collection} with 664\,138 trips (defined as a vehicle moving between two consecutive stops) between 6\,858 bus, metro and ferry stops. The mobile call and Twitter interactions datasets were processed as an undirected temporal network. Public transportation and Air transportation datasets were processed as directed networks with delays (duration of time between departure and arrival) taken into account.

\begin{figure*}[ht!]
    \centering
    \includegraphics[width=\linewidth]{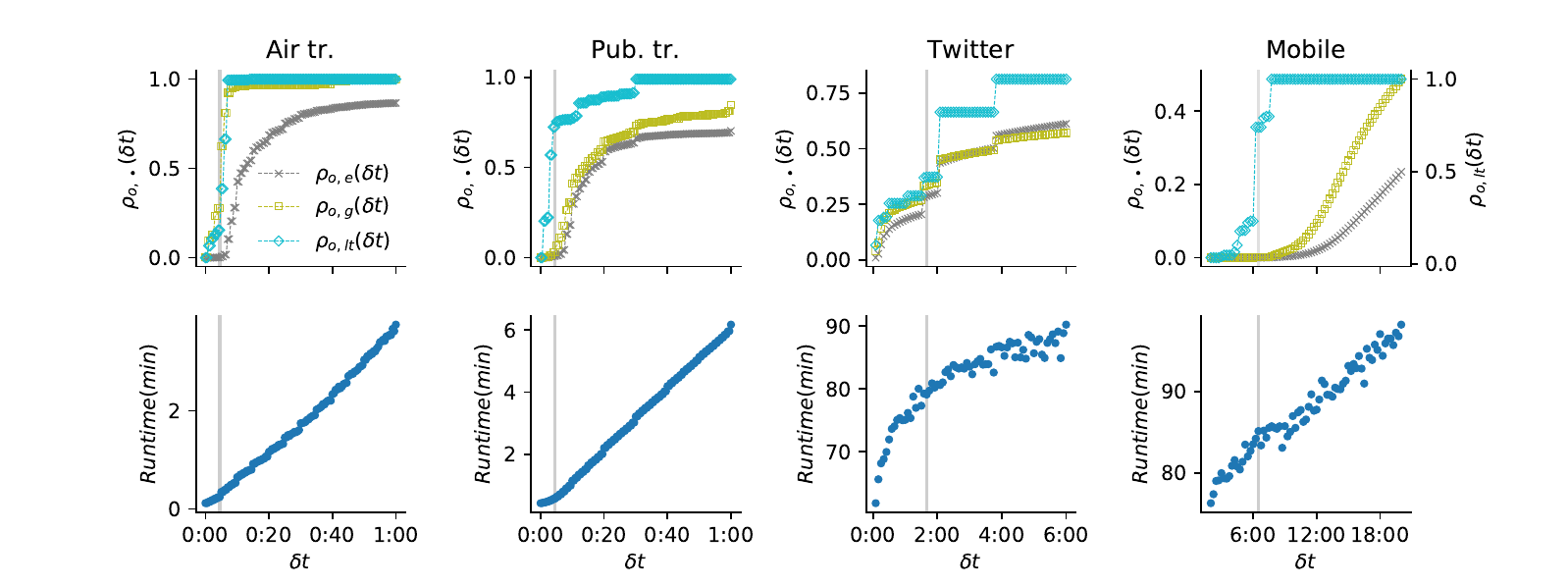}
    \caption{\label{fig:rhos_and_times} Top row displays maximum out-component sizes based on number of events ($\rho_{o,e}$) number of unique nodes ($\rho_{o,g}$) and lifetime of the out-component ($\rho_{o,lt}$). Bottom row shows the median runtime of the algorithm for the estimation of out-component sizes (number of unique events, nodes and lifetime) for different $\delta t$ values. The vertical line corresponds to the $\delta t^*$ value in Table~\ref{tab:real}.}
\end{figure*}

HyperLogLog estimators for Mobile and Twitter networks use $m=2^{10}$ registers. For other datasets $m=2^{14}$ registers were used. Largest out-component sizes were measured with a maximum probability of misidentifying of at most 0.01.

\begin{table}[ht!]
  \begin{center}
\caption{Running times for real-world networks when calculating the reachability (number of unique reachable events, nodes and lifetime) from all events in the network. The $\delta t^*$ corresponds to a waiting time around the time at which there is a jump in the largest out-component size (see the text for details; this corresponds to the vertical line in  Fig.~\ref{fig:rhos_and_times}). As the $\delta t$ values around $\delta t^*$ is of interest for a wide range of studies, the runtime for $\delta t = \delta t^*$ would be representative of the running times for a typical study. The values for $\delta t^*$ are 271 seconds for air and public transport networks, 100 minutes for the Twitter network and 6.5 hours for the mobile network. Baseline running times are measured by calculating out-component size on a sample of 500 events based on Algorithm~\ref{algo:bfs-full} (see Appendix~\ref{sec:bfs}) and extrapolating to estimate running time of exact measurement of out-component size from every event. \emph{Error} column refers to relative standard error for each reachability estimate based on the number of registers $m$ used in HyperLogLog estimator. The times are presented in seconds (s), minutes (m), hours (h), and years (y).
All runtimes are measured using CPU clock time on a mixture of Intel\textsuperscript{\textregistered} Xeon\textsuperscript{\textregistered} E5 2680 v2-4 CPUs (2.40GHz to 2.80GHz) and Gold 6148 (2.40GHz).}
    \label{tab:real}
    \begin{tabular}{lcccccc}
                     & & & \multicolumn{2}{c}{\textbf{Runtime}} & \multicolumn{2}{c}{\textbf{Baseline}}\\
      \textbf{Name} & \textbf{Events} & \textbf{Error} & $\delta t = \infty$ & $\delta t = \delta t^*$ & $\delta t = \infty$ & $\delta t = \delta t^*$\\
      \hline
      Mobile      & 325M     & 3.3\%     & 106m & 85m & 1695y & 21y \\
      Twitter     & 258M     & 3.3\%     & 90m  & 77m & 2409y & 243y \\
      Pub. tr.    & 664K     & 0.81\%    & 59m  & 60s & 19h   & 13m \\
      Air tr.     & 180K     & 0.81\%    & 235m & 17s & 138m  & 60s \\
    \end{tabular}
  \end{center}
\end{table}

Table~\ref{tab:real} provides information on median runtime (as measured by CPU clock time) of the out-component size estimation portion of the implementation. The running time is shown for a $\delta t = \delta t^*$ threshold close to a jump in the largest component size, which is likely to be around the interesting region. We also report the largest possible threshold $\delta t = \infty$ leading to largest running times. For undirected temporal networks (Mobile and Twitter) taking $\delta t$ to infinity does not result in a substantial increase in the running time as most of the increase in the number of event graph links are never considered due to the optimising for redundant links (see Appendix~\ref{sec:redWTEGs}). This, however, is not the case for directed temporal networks as the optimisation method described in Appendix~\ref{sec:redWTEGs} does not apply to directed events. Assuming a homogeneous distribution of events across time, the runtime for event graphs constructed from directed events grows by $\mathcal{O}(\delta t \log \delta t)$ and reaches a maximum at $\delta t = T$ where T is the maximum $\delta t$ between any two events in the network. For the case of instantaneous (non-delayed) events, $T$ is equal to the measurement window of the dataset.

Table~\ref{tab:real} also gives estimates of running times for a breadth-first search type of algorithm for comparison. In these examples, the smallest network with less than $200k$ events takes around the same order of magnitude of time to process with both algorithms. However, even for the second-largest network with around $600k$ events, there is an order of magnitude of difference in the running times. For the larger networks with hundreds of millions of events, the run time jumps down from thousands of years with breadth-first search to order of hours with the new algorithm. This means that large data sets that were previously practically impossible to analyse this way are now accessible even with minimal computational resources.

Figure~\ref{fig:rhos_and_times} shows a more systematic analysis of the running times for the real data, where we vary the $\delta t$ parameter. As previously described in Sec.~\ref{sec:random}, as $\delta t$ is increased larger and larger connected structures begin to form in the event graph. The increase in size is also visible from the largest out-component size curves for the same dataset in Fig.~\ref{fig:rhos_and_times}. This transition period usually marks the most interesting area for further studies. Running times of the breadth-first type of algorithms are in practice dependent on the component sizes and can thus see a dramatic increase in the running times during and after the transition period. Running time plots  (Fig.~\ref{fig:rhos_and_times}) show that as expected our algorithm is not sensitive to these transitions. They show that while for the case of directed networks (air and public transportation) runtime grows almost linearly as a function of $\delta t$, it grows sub-linearly for undirected networks because of the wider range of applied optimisation described in Appendix~\ref{sec:redWTEGs}.

\section{Discussion}

We have presented a method for computing component sizes starting from multiple sources (or reaching multiple destinations) in temporal networks which scales well with the increasing data size. Using simulated networks and real network data we show that the method is efficient enough for us to accurately estimate the $\delta t$-reachability for each of the events in networks with hundreds of millions of events. As a further demonstration of the capabilities of the algorithm, we repeated several results from a previous study~\cite{kivela2018mapping} using accurate estimates for the component sizes instead of using weakly connected components as upper bounds.

Previously temporal network studies have focused on sampling starting points for reachability or simulated spreading processes and exactly calculating statistics based on that sample~\cite{Holme2005Network,holme2015modern}. The sampling approach usually works well for calculating mean values or estimating the parts of distributions where most of the values lie. However, it is not suitable for analysing tails of distributions or extreme values in the networks. Perhaps more importantly, sampling is ill-suited for microscopic analysis of properties of individual nodes or events, which require calculating the reachability from each of them separately. The algorithm presented here is suitable for this type of studies and opens up possibilities for many new kinds of analysis of large data sets.

When presenting the algorithm, we aimed to work at a general level in taking into account various use cases. The definition of a temporal network we used is rather inclusive, although other kinds of hypergraph-type structures could have been considered. Despite these efforts, there are use cases that we did not still consider. Consider, for example, a situation where the edges are available for the paths with some uncertainty such that events $e_1$ and $e_2$ are adjacent with probability $P(e_1, e_2, \delta t)$. In this case the algorithm could be easily used
by simulating many instances of the event graph as result a deterministic random process to measure expected values of in- and out-component sizes. This is important for processes with a stochastic component, such as infection spreading models. Further, we did not discuss multiple sources or targets for the paths. However, as far as we have considered various scenarios such as the above mentioned multiple sources and targets, the algorithms proposed here would have required only minor adjustments. 

Here we have mainly focused on the algorithmic improvements, and used the new method to demonstrate its ability to handle multiple types of networks with sizes varying all the way to hundreds of millions of events. We have barely scratched the surface in the type of analysis, which our new method enables. For example, microscopic network statistics such as centrality measures for nodes could now be defined based on the $\delta t$-reachability counts. Further, theoretical studies of directed temporal percolation in networks are now in our grasp as we can efficiently compute the relevant statistics. Our theoretical and algorithmic contributions allow to study effectively directed percolation phenomena in temporal networks, contrary to earlier works, which are either based on ordered lattices~\cite{hinrichsen2000non} or otherwise unsuitable assumptions for temporal networks. We expect our work on the computational methods to open doors for many future branches of research in data analysis and theory for temporal networks.

\begin{acknowledgments}
The calculations presented in this work were performed using computer resources within the Aalto University School of Science ``Science-IT'' project. Márton Karsai acknowledges support from the ANR project DataRedux (ANR-19-CE46-0008) and the ACADEMICS grant of the IDEXLYON, project of the Université de Lyon, PIA operated by ANR-16-IDEX-0005.
\end{acknowledgments}

\appendix

\section{Implicit construction of weighted temporal event graphs}
\label{sec:event-graph-rep}

To estimate out-component sizes of nodes in event graphs, there is no need to calculate and store the complete static event graph of a temporal network, but it is enough to simply provide fast functions to compute the direct predecessors and successors of any node in $D$. This is accomplished by creating a hash table with each temporal network node in $G$ as the key and a list of all the incident events sorted by time and by the other node of the event as value.

With this setup, to find the direct successors of an event $e$, for each node incident to that event, we look up the list of all incident events from the hash table, find the event $e$ by a binary search and move forward through the list until events have a larger time difference than the maximum allowed waiting time $\delta t$. A union of the two sets of events, extracted for the two ending nodes, would give a complete set of direct successors without pre-calculating the whole event graph. A predecessors function is similarly defined but moving backwards in time in the list of sorted events. Algorithm~\ref{algo:succ} demonstrates the direct successor and predecessor functions for this representation.

\begin{algorithm}[tbp]
    \includegraphics[width=0.90\linewidth]{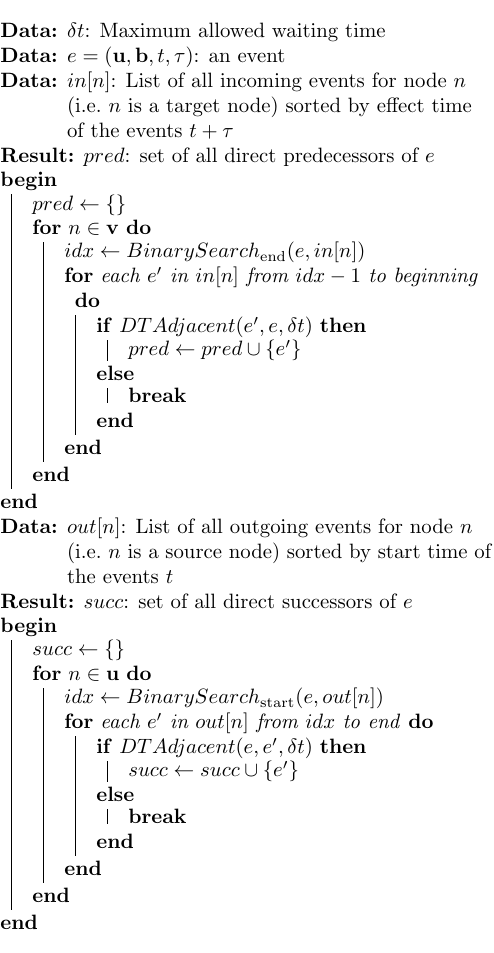}
    \caption{\label{algo:succ} Calculating direct predecessors and successors of an event. $BinarySearch_{\text{start}}(e, vec)$ finds the index of the first event, $e^\prime=(u^\prime,v^\prime,t^\prime,\tau^\prime)$, in $vec$ with start time $t^\prime$ larger or equal to that of the input event $e$. $BinarySearch_{\text{end}}(e, vec)$ similarly finds first event in $vec$ where its ending time $t+\tau$ is not less than that of $e$. Both functions rely on $vec$ being already sorted in ascending order of $t$ or $t+\tau$ respectively. $DTAdjacent(e_1, e_2, \Delta t)$ checks whether $e_2$ is $\delta t$-adjacent to $e_1$.
    }
\end{algorithm}

\section{Locally reduced weighted event graphs}
\label{sec:redWTEGs}

The definition of temporal event graphs allows for certain redundancies by repeating paths between non-directly adjacent events but which are connected via a time respecting path anyway. By removing these redundancies~\cite{mellor2017temporal}, we can accelerate the computation of direct predecessors and successors events while calculating out-components. As an example, let's take an (undirected) event $e_0 = {v_0, u_0, t_0}$ and assume that events $e_1, e_2, e_3\ldots$ at times $t_1, t_2, t_3\ldots$ are the direct successors of $e_0$ through node $v_0$ (i.e., they all share node $v_0$ and $t_0 < t_1 < t_2 < t_3 <\ldots$). However, to represent the connectivity of these events with $e_0$, there is no need to assign $e_2, e_3\ldots$ as direct successors of $e_0$. If we only return $e_1$ as direct successor of $e_0$ through $v_0$, then $e_2$ will be returned as direct successor of $e_1$ through $v_0$, etc., and the obtained out-component would be the same as for the redundant representation. Only if the network allows simultaneous events of the same node (e.g. if $t_1 = t_2$), then we should pay special attention to these cases and assign both $e_1$ and $e_2$ as direct successors of $e_0$ through $v_0$.

\section{Measuring out-component set of a single event}
\label{sec:bfs}

It is trivial to measure the out-component size of a node given the event graph by applying a variant of the breadth-first search algorithm, as demonstrated in Algorithm~\ref{algo:bfs}. This, however, in combination with implicit construction of the event graph as it is described in Appendix~\ref{sec:event-graph-rep}), might not result in the most optimal way to measure out-component size. This is because iterating over direct successors (and predecessors) of a node in the event graph is no longer of linear computational complexity relative to the number of direct successors (or predecessors) of that node.

\begin{algorithm}[tbp]
    \centering
    \includegraphics[width=0.90\linewidth]{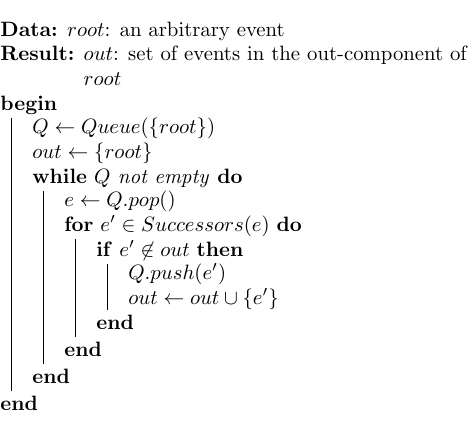}
    \caption{\label{algo:bfs} Calculating exact out-component of an events from the static event graph representation described in Appendix~\ref{sec:event-graph-rep}.}
\end{algorithm}

\begin{algorithm}[tbp]
    \centering
    \includegraphics[width=0.90\linewidth]{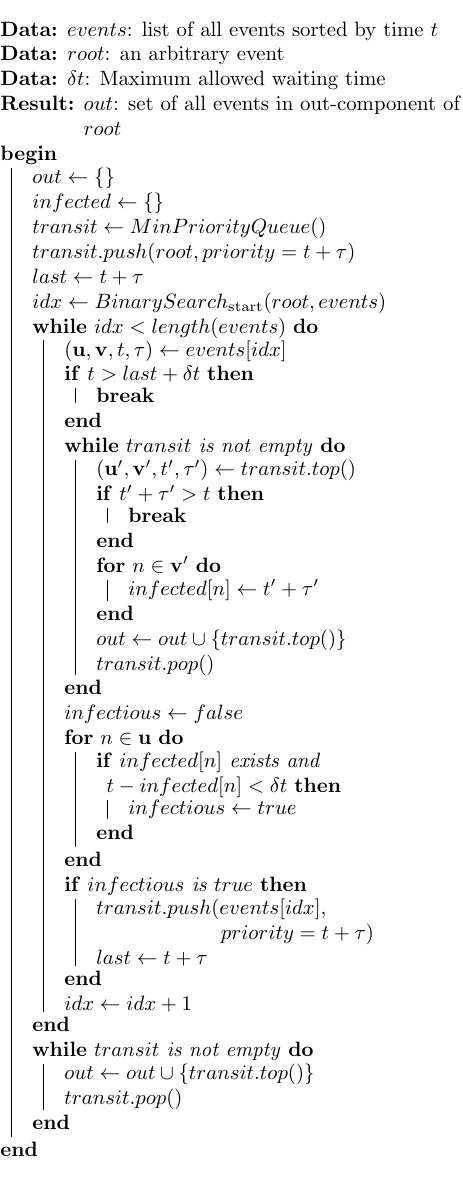}
    \caption{\label{algo:bfs-full} Calculating exact out-component of an events without explicitly forming an event graph. Contrary to Algorithm~\ref{algo:bfs} the complexity of this algorithm does not depend on representation of static event graph and therefore only requires one scan through the list of events. For the case of delayed connections, it also needs to track all the events that have been initiated but are not concluded in a minimum priority queue data structure.}
\end{algorithm}

Algorithm~\ref{algo:bfs-full} describes a method for calculating out-component size by scanning through the list of events once. For the case of non-delayed events, regardless of directedness, it is possible to dispense with priority queue and provide a much simpler implementation and a computational complexity of $\mathcal{O}(|E|)$ where $|E|$ denotes the number of events. For the case of delayed events, however, computational complexity will depend on the number of simultaneous \emph{in transit} events and the selected implementation of the priority queue.

\bibliography{references}
\end{document}